\documentclass[aip, jmp, preprint]{revtex4-1}
\usepackage{amsmath}
\usepackage{amssymb}
\usepackage{graphicx}
\usepackage{bbold}

\newtheorem{theorem}{Theorem}[section]

\newenvironment{proof}[1][Proof]{\begin{trivlist}
\item[\hskip \labelsep {\bfseries #1}]}{\end{trivlist}}

\newcommand{\qed}{\nobreak \ifvmode \relax \else
      \ifdim\lastskip<1.5em \hskip-\lastskip
      \hskip1.5em plus0em minus0.5em \fi \nobreak
      \vrule height0.75em width0.5em depth0.25em\fi}

\begin{document}


\title{A completeness-like relation for Bessel functions} 



\author{Paulo H. F. Reimberg}

\affiliation{Sorbonne Universit\'es, UPMC Univ Paris 6 et CNRS, UMR 7095, Institut d'Astrophysique de Paris, 98 bis bd Arago, 75014 Paris, France}
\affiliation{CEA - CNRS, UMR 3681, Institut de Physique Th\'eorique, F-91191 Gif-sur-Yvette, France}
\email[E-mail address: ]{paulo.flose-reimberg@cea.fr}
\author{L. Raul Abramo}
\affiliation{Instituto de F\'isica, Universidade de S\~ao Paulo, 
CP 66318, 05314-970, S\~ao Paulo, Brazil}



\begin{abstract}
Completeness relations are associated through Mercer's theorem to complete orthonormal basis of square integrable functions, and prescribe how a Dirac delta function can be decomposed into basis of eigenfunctions of a Sturm-Liouville problem. We use Gegenbauer's addition theorem to prove a relation very close to a completeness relation, but for a set of Bessel functions not known to form a complete basis in $L^2[0, 1]$.
\end{abstract}


\maketitle 

\section{Introduction}

If one studies Helmholtz equation on the interior of a sphere of unitary radius subjected to condition of regularity of the solution at $r=0$, and such that the solution vanishes at $r=1$, the Green's function can be constructed in terms of the eigenfunctions of the Sturm-Liouville problem using Mercer's formula, from which follows the \emph{completeness relation} for spherical Bessel functions:

\begin{equation}
\label{completeness_j}
\sum_{n=1}^{\infty} \frac{j_l(a_n^l x) j_l(a_n^l y)}{[j_{l+1}(a_n^l)]^2} = \frac{1}{2} \delta(x - y)
\end{equation}
where $0 \leq x, y \leq 1$, and $a_n^l$ is the n-th zero of the spherical Bessel function of order $l$. The set of functions $j_l(a_n^l x)$ forms a complete basis of orthogonal functions on $L^2[0, 1]$, and this kind of relations are very useful in quantum mechanics, for example. 

Since Bessel functions are associated to representations of Euclidean group, they appear abundantly in physical problems, and in particular on the problem of random flights, that are the random motions in a $D$-dimensional euclidean space performed by a particle that, always with constant speed, change the direction of its motion after a set of instants of time distributed accordingly to some law \cite{hughes, dutka, orsinger, degregorio_12, kolesnik_06, kolesnik_08, caer_10, caer_11, pogorui_11}. The probability for the walker to be at a distance $r$ from the origin of the motion after $n$ changes of directions is related to an integral of $n+1$ Bessel functions \cite{watson}. We can study random flights that start at a subspace $D_1$-dimensional of a $D_2$-dimensional space and, after a given number of steps on the space of smaller dimensions, accesses the space of larger dimension. Such problems can be easily imagined in physical situations and is particularly realized in Cosmic Microwave Background (CMB) physics \cite{reimberg2015random, cmb_flight}. The completeness-like relation that we shall prove here plays the role of a consistency relation for the decomposition of such flights, and was first found in the context of CMB physics \cite{cmb_box}.

\section{Completeness-like relations for Bessel functions}
\label{compl_like}

\begin{theorem}[Completeness-like relation for Bessel functions]
\label{completeness_like_theorem}
Let $0 \leq x , y \leq 1$, $D_2 \geq D_1$, $D_2 - D_1 > 0$ an even number, and $\lambda_k^{D_2/2-1}$ the k-th zero of the Bessel function $J_{D_2/2-1}( \, . \, )$. Then:
\begin{equation}
\label{completeness_like_th}
\sum_{k=1}^{\infty} \frac{J_{D_1/2-1} \left( \lambda_k^{D_2/2-1} x \right)  J_{D_1/2-1} \left( \lambda_k^{D_2/2-1} y \right) }{ J_{D_2/2}^2 \left( \lambda_k^{D_2/2-1} \right)}  = 
\frac{1}{2 x} \delta \left( x - y \right) \, .
\end{equation}
\end{theorem}

\begin{proof}
In order to establish our result, we shall look at the integral
\begin{equation}
\label{int_three}
\mathcal{I} := \int dq \, q^{D_2-1} \, \frac{J_{D_2/2-1}(qr)}{(qr)^{D_2/2-1}} \, \frac{J_{D_2/2-1}(qs_1)}{(qs_1)^{D_2/2-1}} \, \frac{J_{D_1/2-1}(q r_1)}{(q r_1)^{D_1/2-1}} \, 
\end{equation}
in three different ways. First, we can contract the two Bessel functions of order $D_2/2-1$ using Eq. \eqref{contraction_diff_dim}, what yields:

\begin{eqnarray}
\label{contr_three}
\mathcal{I} & = & \frac{2^{D_1/2-1}}{(r s_1)^{\Delta/2}} \frac{ (\Delta/2)! \Gamma(D_1/2-1)}{2 \pi  \Gamma(D_1-2+\Delta/2)} \int_0^{\pi} d \alpha \sin^{D_1-2} \alpha \, 
C_{\Delta/2}^{D_1/2-1} (\cos \alpha) \nonumber\\ & & \times
\frac{1}{(\rho r_1)^{D_1/2-1}} \int dq \, q \, J_{D_1/2-1} (q \rho) \, J_{D_1/2-1}(q r_1) \, ,
\end{eqnarray}
with $\rho^2 := r^2 + s_1^2 - 2 r s_1 \cos \alpha$, and
\begin{equation}
\label{orthog_J}
\int dq \, q \, J_{D_1/2-1} (q \rho) \,  J_{D_1/2-1}(q r_1) = \frac{1}{r_1} \delta( \rho - r_1) \, .
\end{equation}

Secondly, we can regard the integral $\int dq \, q \, J_{D_1/2-1} (q \rho) \, J_{D_1/2-1}(q r_1)$ as a function of $\rho$, with $r_1$ fixed. Since Eq. \eqref{int_three} is a discontinuous Weber-Schafheitlin integral \cite{watson}, we know that it vanishes identically if $r_1, s_1$, and $r$ do not form a triangle. We must have, therefore, $|s_1 - r| \leq \rho \leq s_1+r =: S$. Since the domain of $\rho$ is contained in $[0, S]$, we can construct the Fourier-Bessel decomposition \cite{watson, hochstadt}:

\begin{equation}
\label{FB_zwei}
\int dq \, q \, J_{D_1/2-1} (q \rho) \, J_{D_1/2-1}(q r_1) = \sum_{k=1}^{\infty}
c_k J_{D_1/2-1} \left( \lambda_k^{D_1/2-1} \frac{\rho}{S} \right)
\end{equation}
with
\begin{equation}
\label{FB_coeff}
c_k = \frac{2}{S^2 J_{D/2}^2 \left(\lambda_k^{D/2-1} \right)} \int_0^{S} d \rho \, \rho 
\int dq \, q \,  J_{D_1/2-1} (q \rho) \, J_{D_1/2-1}(q r_1) \, 
J_{D/2-1} \left( \lambda_k^{D/2-1} \frac{r}{S} \right) \, ,
\end{equation}
and $\lambda_k^{\nu}$ being the $k$-th zero of the Bessel function $J_{\nu} ( \, \cdot \, )$.
The integral $\int dq \, q \, J_{D_1/2-1} (q \rho) \, J_{D_1/2-1}(q r_1)$
vanishes identically if $\rho > S$, what allows us to extend the limit of integration  to infinity. Applying the Fourier-Bessel integral \cite{whittaker_watson}
\begin{equation}
f(x) = \int_0^{\infty} J_n(tx) \, t \left[ \int_0^{\infty} f(x') J_n(tx') x' dx' \right] dt
\end{equation}
to Eq. \eqref{FB_coeff}, and inserting the result into Eq. \eqref{FB_zwei},
we obtain:
\begin{eqnarray}
\label{FB_zwei_final}
\int dq \, q \, J_{D_1/2-1} (q \rho) \,  J_{D_1/2-1}(q r_1) & = & \sum_{k=1}^{\infty} \frac{2}{S^2 J_{D_1/2}^2 \left( \lambda_k^{D_1/2-1} \right)}  \nonumber\\ & & \times
J_{D_1/2-1} \left( \lambda_k^{D_1/2-1} \frac{\rho}{S} \right)  J_{D_1/2-1} \left( \lambda_k^{D_1/2-1} \frac{r_1}{S} \right) \, .
\end{eqnarray}
Collecting the partial results presented in Eqs. \eqref{contr_three}, and \eqref{FB_zwei_final}, we write integral \eqref{int_three} as:

\begin{eqnarray}
\label{contr_1}
\mathcal{I} & = &  \frac{2^{D_1/2-1}}{(r s_1)^{\Delta/2}}  \frac{ (\Delta/2)! \Gamma(D_1/2-1)}{2 \pi \Gamma(D_1-2+\Delta/2)}  \int_0^{\pi} d \alpha \sin^{D_1-2} \alpha \,
C_{\Delta/2}^{D_1/2-1} (\cos \alpha)  \nonumber\\ & & \times  \frac{1}{(\rho r_1)^{D_1/2-1}}
\sum_{k=1}^{\infty} \frac{2}{S^2 J_{D_1/2}^2 \left( \lambda_k^{D_1/2-1} \right)}   J_{D_1/2-1} \left( \lambda_k^{D_1/2-1} \frac{\rho}{S} \right)  
 J_{D_1/2-1} \left( \lambda_k^{D_1/2-1} \frac{r_1}{S} \right) \, .
\end{eqnarray}

We can now consider the third way of treating Eq. \eqref{int_three}. 
Since $r, r_1$, and $s_1$ must be related to sides of a triangle, the inequality $|r_1 - s_1| \leq r \leq r_1+s_1 := \tilde{S}$ must hold. Hence the integral $\int dq \, q \, J_{D_2/2-1}(qr) \, J_{D_2/2-1}(qs_1) \, \frac{J_{D_1/2-1}(q r_1)}{(q r_1)^{D_1/2-1}}$ can be expressed in terms of a Fourier-Bessel series:

\begin{eqnarray}
& & \int dq \, q \,  J_{D_2/2-1}(qr) \, J_{D_2/2-1}(qs_1) \, \frac{J_{D_1/2-1}(q r_1)}{(q r_1)^{D_1/2-1}}   =  \sum_{k=1}^{\infty} \frac{2}{\tilde{S}^2 J_{D_2/2}^2 \left( \lambda_k^{D_2/2-1} \right)}
\nonumber\\ & & \times
J_{D_2/2-1} \left( \lambda_k^{D_2/2-1} \frac{r}{\tilde{S}} \right) 
J_{D_2/2-1} \left( \lambda_k^{D_2/2-1} \frac{s_1}{\tilde{S}} \right) 
\frac{ J_{D_1/2-1} \left( \lambda_k^{D_2/2-1} \frac{r_1}{\tilde{S}} \right)}{ \left( \lambda_k^{D_2/2-1} \frac{r_1}{\tilde{S}} \right)^{D_1/2-1}} \, .
\end{eqnarray}
We can use Eq. \eqref{contraction_diff_dim} to contract the two Bessel functions of order $D_2/2-1$, obtaining:

\begin{eqnarray}
\label{contr_2}
\mathcal{I} & = &  \frac{2^{D_1/2-1}}{(r s_1)^{\Delta/2}} \, \frac{ (\Delta/2)! \Gamma(D_1/2-1)}{2 \pi \Gamma(D_1-2+\Delta/2)}  \int_0^{\pi} d \alpha \sin^{D_1-2} \alpha \,
C_{\Delta/2}^{D_1/2-1} (\cos \alpha) \nonumber\\ & & \times \frac{1}{(\rho r_1)^{D_1/2-1}}
\sum_{k=1}^{\infty} \frac{2}{\tilde{S}^2 J_{D_2/2}^2 \left( \lambda_k^{D_2/2-1} \right)}   J_{D_1/2-1} \left( \lambda_k^{D_2/2-1} \frac{\rho}{\tilde{S}} \right)   J_{D_1/2-1} \left( \lambda_k^{D_2/2-1} \frac{r_1}{\tilde{S}} \right) \, ,
\end{eqnarray}
where $\rho$ is also given by the relation $\rho^2 = r^2 + s_1^2 - 2 r s_1 \cos \alpha$.

Since the function $\frac{\sin^{D_1 -2} \alpha \; C_{\Delta/2}^{D_1/2-1} (\cos \alpha)}{\rho^{D_1/2-1}}$ does not vanish for all $\alpha \in [0, \pi]$ for any $D_1, \Delta$, we conclude, comparing  Eqs. \eqref{contr_1} and \eqref{contr_2}, that:
\begin{eqnarray}
& & \sum_{k=1}^{\infty} \frac{2}{S^2 J_{D_1/2}^2 \left( \lambda_k^{D_1/2-1} \right)}  J_{D_1/2-1} \left( \lambda_k^{D_1/2-1} \frac{\rho}{S} \right)  J_{D_1/2-1} \left( \lambda_k^{D_1/2-1} \frac{r_1}{S} \right) \nonumber\\ & &  = \sum_{k=1}^{\infty} \frac{2}{\tilde{S}^2 J_{D_2/2}^2 \left( \lambda_k^{D_2/2-1} \right)}  J_{D_1/2-1} \left( \lambda_k^{D_2/2-1} \frac{\rho}{\tilde{S}} \right)  J_{D_1/2-1} \left( \lambda_k^{D_2/2-1} \frac{r_1}{\tilde{S}} \right) \, .
\end{eqnarray}
The left hand side of this equation is equal to $\frac{1}{r_1} \delta(\rho - r_1)$ because of the orthogonality of Bessel functions, as stated in Eq. \eqref{orthog_J}, and therefore

\begin{equation}
\label{completeness_like}
\sum_{k=1}^{\infty} \frac{ J_{D_1/2-1} \left( \lambda_k^{D_2/2-1} \frac{\rho}{\tilde{S}} \right)  J_{D_1/2-1} \left( \lambda_k^{D_2/2-1} \frac{r_1}{\tilde{S}} \right)}{ J_{D_2/2}^2 \left( \lambda_k^{D_2/2-1} \right)}  = \frac{1}{2 \left( \frac{r_1}{\tilde{S}} \right)} 
\delta \left( \frac{\rho}{\tilde{S}} - \frac{r_1}{\tilde{S}} \right) \, ,
\end{equation}
as we wanted to demonstrate. \qed
\end{proof}

The form of the equation Eq. \eqref{completeness_like} resembles the well known completeness relation \eqref{completeness_j} with the striking difference that in Eq. \eqref{completeness_j} the zeros of the Bessel function appearing in the sum belong the to the Bessel function of that order. In Eq. \eqref{completeness_like}, however, we sum over Bessel functions of order $D_1/2-1$ with zeroes in their argument belonging to the Bessel function of order $D_2/2-1$. 

The integral $\mathcal{I}$ is directly associated, on the theory of random flights, to the probability density of finding the walker at a distance $r$ from the origin after realizing one step of length $r_1$ in a space of dimension $D_1$, and one step of length $s_1$ on a $D_2$-dimensional space. These two steps can actually be effective steps associated to contraction of all steps executed on the spaces of each dimension \cite{reimberg2015random}. The completeness-like relation proved here assures that contraction of steps in any possible way leads to consistent results.

\section{Discussion}

We have demonstrated an identity formally close to the completeness relation for Bessel functions, that we called completeness-like relation. The curious fact about this relation is that sequences of products of Bessel functions evaluated at the zeroes of a Bessel function of different order converges to a Dirac distribution. This may constitute an step into the generalization of this result in the context of Schl\"omilch's series.

The identity \eqref{completeness_like} is direct consequence of the Gegenbauer addition theorem and  Fourier-Bessel series expansions, and holds for all $D_1, D_2 > 0$ such that $D_2 - D_1 = \Delta \in \mathbb{N}$, $\Delta$ even. $D_1, D_2 > 0$ must be positive because the set $\{ \sqrt{x} J_{\nu} ( \lambda_k^{\nu} x) \}_{k \in \mathbb{N}}$ only forms a complete set in $L^2(0, 1)$ for $\nu > -1$, which is necessary for definiteness of the Fourier-Bessel series.

A theorem demonstrated in Ref. \onlinecite{boas} states:
\emph{ If $\nu > -1/2$, the set $\{x^{1/2} J_{\nu}(\lambda_n x) \}$ forms a complete 
(hence total) sequence in $L^p(0, 1)$, $1 \leq p < \infty$, if for all sufficiently large 
$n$ we have:
\begin{displaymath}
0 < \lambda_n \leq \pi \left(n + \frac{1}{4} + \frac{\nu}{2} -\frac{1}{2p} \right) \; .
\end{displaymath} }
Completeness, in the sense of this theorem, means that if 
$\int_0^1 x^{1/2} J_{\nu}(\lambda_n \, x) g(x) dx =0$ for all $\nu > -1/2$, 
with $g(x) \in L^p[0, 1]$, then $g(x)$ vanishes almost everywhere on $(0, 1)$ \cite{higgins}. 
For large values of the argument, the Bessel functions $J_{\nu}(z)$ 
behave like $\cos(z -\nu \pi/2 - \pi/4)$ \cite{watson}, and the theorem 
assures that the set $\{x^{1/2} J_{\nu}(\lambda_n x) \}$ is complete in the Hilbert space $L^2[0, 1]$ when $\lambda_n$ are taken to be the positive real zeros of  $J_{\mu}(z)$, but only if 
$\mu \leq \nu$. Since in our case $D_2\geq D_1$ (which corresponds to $\mu \geq \nu$), 
what we have shown is a completeness-like relation for a set of functions 
whose completeness cannot be discussed inside the scope of the aforementioned theorem. 
It is not known to the authors whether there are extensions of this theorem that 
include the case presented here, nor is it clear what is the meaning of the 
completeness-like relation that we obtained in the general context of the 
completeness of sets of Bessel functions in $L^p[0, 1]$.

\begin{acknowledgments}

The authors would like to thank Jo\~ao C. A. Barata. This work was supported by FAPESP. PR also thanks the Agence Nationale de la Recherche under the grant ANR-12-BS05-0002 for financial support.
\end{acknowledgments}

\appendix

\section{Gegenbauer addition theorem}
For Bessel functions of first kind, the Gegenbauer addition theorem states that \cite{watson}:

\begin{equation}
\label{gegenbauer_watson}
\int_0^{\pi} \frac{J_{\nu}(Z^2 + z^2 - 2 z Z \cos \alpha)}{(Z^2 + z^2 - 2 z Z \cos \alpha)^{\nu/2}} \, 
C_n^{\nu}(\cos \alpha) \, \sin^{2 \nu} \alpha \, d \alpha = 
\frac{\pi \, \Gamma(2 \nu +n)}{2^{\nu-1} \, n! \, 
\Gamma(\nu)} \, \frac{J_{\nu+n}(Z)}{Z^{\nu}} \, \frac{J_{\nu+n}(z)}{z^{\nu}} \, 
\end{equation}
where $\nu > -1/2$, $\nu \in \mathbb{R}$, $m \in \mathbb{N}$, and $C_n^{\nu}(\cos \alpha)$ are Gegenbauer polynomials, defined by the
relation:

\begin{displaymath}
\frac{1}{(1 - 2 t \cos \alpha  + t^2)^{\nu}} = \sum_{n=0}^{\infty}
C_n^{\nu} (\cos \alpha) \, t^n \, .
\end{displaymath}
As immediate consequence we can construct the contraction of two Bessel function of a given order into a Bessel function of smaller order:

\begin{eqnarray}
\label{contraction_diff_dim}
\frac{J_{D_2/2-1} (q l_1)}{\left( q l_1 \right)^{D_2/2-1}} 
\frac{J_{D_2/2-1} (q l_2)}{\left( \frac{q l_2}{2} \right)^{D_2/2-1}} & = &
\frac{2^{D_1/2-1}}{q^{\Delta} \left( \frac{l_1 l_2}{2} \right)^{\Delta/2}}  \, 
\frac{ \left(\Delta/2 \right)! \Gamma(D_1/2 -1)}{2 \pi \Gamma(D_1 -2 + \Delta/2)}
\nonumber\\ & & \times
\int_0^{\pi} d \alpha \, \frac{J_{D_1/2-1} (q \rho)}{\left( \frac{q \rho}{2} \right)^{D_1/2-1}} \,
C^{D_1/2-1}_{\Delta/2} (\cos \alpha) \, \sin^{D_1 - 2} \alpha  \, ,
\end{eqnarray}
where $D_2 = D_1 + \Delta$, $\Delta \in \mathbb{N}$ is an even number, and $\rho^2 = l_1^2 + l_2^2 -2 l_1 l_2 \cos \alpha$. 

Here we see that the Gegenbauer addition theorem allows us to contract the product of two Bessel functions of order $D_2/2-1$ into a marginalization over one Bessel function of order $D_1/2-1$ \emph{if} $\Delta$ is an even number, since the lower index of the Gegenbauer polynomials must be an integer. Even if the Gegenbauer polynomials can be defined (through its hypergeometrical representation) for non-integer lower indexes, extensions of the Gegenbauer addition theorem to these cases are not known by the authors.

\bibliography{referencias}



%
%

%



\end{document}